\documentclass[10pt, conference, compsocconf]{IEEEtran}
\IEEEoverridecommandlockouts

\usepackage{amsmath,amssymb,amsfonts}
\usepackage{textcomp}
\usepackage{booktabs}
\usepackage[table]{xcolor}
\usepackage{multirow}
\usepackage{url}
\usepackage{siunitx}
\usepackage{hyperref}

\usepackage{framed}

\usepackage{graphicx} 
\usepackage[T1]{fontenc}    
\usepackage{hyperref}       
\usepackage{url}            
\usepackage{booktabs}       
\usepackage{amsfonts}       
\usepackage{nicefrac}       
\usepackage{microtype}      
\usepackage{graphicx}
\usepackage[backend=biber,style=numeric]{biblatex}

\graphicspath{ {./imgs/} }

\usepackage{algorithm} 
\usepackage{algpseudocode} 
\usepackage{amsmath}

\title{Generating Minimalist Adversarial Perturbations to Test Object-Detection Models: An Adaptive  Multi-Metric Evolutionary Search Approach}

\author{\IEEEauthorblockN{Cristopher McIntyre-Garcia, Adrien Heymans,
Beril Borali, Won-Sook Lee and
Shiva Nejati}
\IEEEauthorblockA{\{cmcin019,aheym065,bbora017,wslee,snejati\}@uottawa.ca \\
University of Ottawa, Ottawa, Canada}}

\begin{document}

\maketitle

\begin{abstract}
Deep Learning (DL) models excel in computer vision tasks but can be susceptible to adversarial examples. This paper introduces Triple-Metric EvoAttack (TM-EVO), an efficient algorithm for evaluating the robustness of object-detection DL models against adversarial attacks. TM-EVO utilizes a multi-metric fitness function to guide an evolutionary search efficiently in creating effective adversarial test inputs with minimal perturbations. We evaluate TM-EVO on widely-used object-detection DL models, DETR and Faster R-CNN, and open-source datasets, COCO and KITTI. Our findings reveal that TM-EVO outperforms the state-of-the-art EvoAttack baseline, leading to adversarial tests with less noise while maintaining efficiency.   
\end{abstract}


\section{Introduction}\label{Section Introduction}

Recent advancements in Deep Learning (DL) models have led to remarkable progress in computer vision tasks~\cite{lin2022survey}.  DL-based computer vision models, despite their advanced capabilities, are vulnerable to adversarial attacks~\cite{goodfellow2014explaining}, which are meticulously designed inputs that exploit the model's weaknesses to cause it to make errors. These attacks often involve subtly altered data indistinguishable to humans but lead the model to incorrect predictions.  This paper specifically focuses on generating challenging adversarial attacks to assess the robustness of object-detection DL models.


Several research efforts have focused on generating adversarial tests for object-detection models using white-box and black-box algorithms~\cite{brendel2019accurate,alzantot2019genattack,bartlett2023strengths,chan2022evoattack,mi2023adversarial}. White-box approaches require access to the model's architecture and weights, which may not always be available, for example,  due to security concerns. In contrast, black-box algorithms are more suitable in these situations~\cite{bartlett2023strengths}. These black-box approaches typically use search-based testing algorithms~\cite{alzantot2019genattack,bartlett2023strengths,chan2022evoattack}. For example, EvoAttack~\cite{chan2022evoattack} introduces a generational Genetic Algorithm (GA) for creating adversarial attacks on object-detection models. While EvoAttack demonstrates the effectiveness of search-based testing in creating attacks, it has not specified if these attacks are optimized for minimal perturbations nor explored the possibility of creating subtler attacks with fewer image modifications for object-detection models.


This paper proposes Triple-Metric EvoAttack (TM-EVO), an approach based on GA that generates adversarial tests for object-detection models. TM-EVO enhances EvoAttack by introducing \emph{an adaptive multi-metric} fitness measure. This measure not only facilitates the generation of attacks but minimizes noise interference in the generated attacks while maintaining efficiency. Specifically, the plateau-based adaptation technique of TM-EVO dynamically updates the weights used to combine the metrics into TM-EVO's fitness function. In addition, it adapts the mutation and the noise reduction operators dynamically based on the search progress. This dynamic adaptation seeks to balance the likelihood of detecting attacks, i.e.,  test input images that mislead the model under test, while ensuring that the level of noise added to the images remains minimal.

We evaluate TM-EVO  on two object-detection models, DETR~\cite{carion2020end} and Faster R-CNN~\cite{ren2015faster}, and using two open-source datasets, COCO~\cite{lin2015microsoft} and KITTI~\cite{geiger2013vision}. Our results show that TM-EVO outperforms the state-of-the-art EvoAttack baseline in attack generation, introducing, on average, 60\% less noise, as measured by the $L_0$ norm metric. This metric is one of the two commonly used to assess the amount of noise introduced in adversarial images~\cite{brendel2019accurate}. As for the $L_2$ norm, the other metric for measuring noise reduction, TM-EVO still outperforms the baseline on average, though by a smaller margin. The improvement in noise reduction by TM-EVO over the baseline does not result in higher run times, as the overall average running time of both approaches remains the same. 

\textbf{Contributions.} (1) A triple-metric fitness function with adaptive weights to balance the trade-off between the number and degree of perturbations and the efficiency in developing successful attacks. (2) An adaptive noise reduction mechanism to reduce ineffective perturbations, achieving more optimal noise levels in successful attacks. (3) An empirical evaluation for   Convolutional Neural Networks (CNN)-based (Faster R-CNN) and Transformer-based (DETR) object detection models.


\section{Related Work}
Brendel et al. \cite{brendel2019accurate} propose gradient-based algorithms to assess model robustness by generating adversarial examples and measuring the time and noise needed for successful attacks. However, these gradient-based algorithms require white-box models, needing access to the network's architecture and weights, which may not be available.

GenAttack ~\cite{alzantot2019genattack} uses a genetic algorithm for generating black-box adversarial attacks on classification models.  Bartlett et. al.~\cite{bartlett2023strengths} extends GenAttack by introducing a multi-objective algorithm for adversarial testing of classification models. This new algorithm aims to minimize noise in adversarial examples and includes a noise removal feature.  Similarly, EvoAttack~\cite{chan2022evoattack} builds upon GenAttack, adapting it for adversarial attacks on object-detection models. This adaptation uses the aggregation of confidence scores from detected objects in an image as a fitness function. EvoAttack introduces noise in areas where objects are detected and uses an adaptive procedure to gradually add noise during the search iterations. Our work is similar to EvoAttack in performing adversarial attacks on object-detection models. However, we enhance this approach by introducing a multi-metric fitness measure that evaluates detection evasion and perturbation minimization without sacrificing run time efficiency. The noise reduction measure is complemented by our adaptive fitness function. This function dynamically balances confidence scores that capture the effectiveness of attacks with noise reduction.

\section{Adversarial Attack Generation}\label{Section Approach}
Algorithm~\ref{alg:tmevo} shows TM-EVO in detail.
TM-EVO essentially implements a Genetic algorithm for adversarial attack generation. Briefly, the algorithm begins with an initial population of individuals, each formed by introducing minor mutations to a given input image, i.e., the original image. It then evolves this population iteratively, selecting promising individuals, combining their traits through crossover, and adding mutations to maintain diversity. This algorithm repeats until either an attack occurs -- when the object-detection model under test (MUT) fails to recognize any object in a generated image -- or the maximum number of generations is reached.

The key novel components of TM-EVO include: (1)~its multi-metric fitness function (Line~6 of Algorithm~\ref{alg:tmevo}), (2)~the plateau-based adaptation applied after fitness computation (Lines~7-9 of Algorithm~\ref{alg:tmevo}), and (3)~the noise reduction mechanism applied after the mutation operator (Line~20 of Algorithm~\ref{alg:tmevo}). Below, we present the novel components of Algorithm~\ref{alg:tmevo} after discussing the attack types adopted in our work. 

\begin{algorithm}
    \caption{Triple-Metric EvoAttack (TM-EVO)}\label{alg:tmevo}
    \scriptsize
    \textbf{Input} $I_{\mathit{ori}}$: original image; $N$: population size; $G$: generations; $\delta$: degree of perturbation; $\rho$: mutation rate; $\bar{\rho}$: noise reduction probability; $w = (w_1, w_2, w_3)$: metric weight(s); $G_p$: Number of generations for adaptation plateau;
    \newline
    \textbf{Result} Adversarial attack image
    \begin{algorithmic}[1]
    \State $P \gets \emptyset$
    \For{$i=1 \to N $}
    \Comment Initialize population
        \State $P \gets P \cup \{I_{\text{ori}} + U(-\delta_{\text{max}} , \delta_{\text{max}}) \}$
    \EndFor
    \For{$g=1 \to G$}
        \State $\mathit{Fitnesses} \gets \text{ComputeFitness}(P, w)$
        \If{\text{IsPlateau($G_p$)}} 
        \Comment Check plateau status
            \State Update($w, \bar{\rho}$)
            \Comment Plateau-based adaptation of weights
        \EndIf
        \State $I \gets$ \text{Image with lowest (best) fitness in} $P$
        \Comment Find  best member
        \If{$\text{isAttack}(I)$}
                    \Comment  MUT detects no objects above a $0.9$ confidence in $I$
            \State \Return $I$

        \EndIf
        \State $\mathit{Best} \gets \{I\}$
         \State $Q \gets \emptyset$
        \For{$i=2 \to N$}
            \State $p_1, p_2$ $\gets$ SampleParents($P$) 
            \State $child \gets \text{Crossover}(p_1, p_2)$
            \State $child_{\text{mut}} \gets \text{AdaptiveMutation}(child, \rho, \delta)$
            \State $child_{\text{mut}} \gets \text{MutationReduction}(child_{\text{mut}}, \bar{\rho})$
            \Comment Mutation noise reduction
            \State $Q \gets Q \cup \{child_{\text{mut}}\}$
        \EndFor
         \State $P \gets Q$
    \EndFor
    \State \Return $\mathit{Best}$
    \end{algorithmic}
\end{algorithm}

\subsection{Attack Type}
The primary goal of an object-detection model is to assign each object in a given image a label from a predefined set of labels and draw a bounding box around the object. Each object in the input image is assigned a confidence score which denotes the probability that the object exists and is properly represented by its label and bounding box. Only objects whose confidence scores exceed a certain threshold are considered to be detected. Two types of strategies exist for performing adversarial attacks: \textit{Targeted} attacks that aim to change object labels to a specific, predetermined label; and \textit{untargeted} attacks that seek to avoid detection entirely. An attack is deemed successful when a MUT fails to detect objects in an image that are normally identifiable by humans. TM-EVO focuses on \textit{untargeted} attacks: it is therefore our intention to generate mutated images whose objects are not detectable by the MUT. 

\subsection{Multi-Metric Fitness and Plateau-Based Adaptation}
\label{subsec:fit}
Similar to  EvoAttack \cite{chan2022evoattack}, TM-EVO's fitness function leverages the confidence scores generated by the object-detection MUT for each object in a given image. However, unlike EvoAttack, we do not solely rely on the MUT's confidence scores to compute fitness values for the individual images generated during the search. Instead, we use a hybrid function combining three different metrics defined below. 

Let $I_{\mathit{ori}}$ be the original image we aim to perturb  to generate an attack for the MUT.  Let $I$ be an image, i.e., candidate solution, in the search population. TM-EVO computes the fitness value for $I$  based on the following three metrics: 

(1) $M_1$ computes  the average confidence scores of all objects detected in $I$ by the MUT: 

{\small\begin{equation}\label{objective_1_equation}
    M_1(I) = \frac{\sum_{i=0}^{n} conf_i }{n}
\end{equation}}
where $n$ represents the number of objects in $I_{\mathit{ori}}$, and ${conf_i}$ denotes the confidence score of the ${i}^{th}$ object, as calculated by the MUT on $I$. Lowering $M_1$ raises the likelihood that $I$ represents an attack, meaning objects on $I$ may go undetected or be misclassified.

(2)~$M_2$ computes  the number of pixels modified in $I$ compared to $I_{\mathit{ori}}$ divided by their total number of pixels:

{\small\begin{equation}\label{objective_2_equation}
    M_2(I) = \frac{pixel_{mut}}{pixel_{total}}
\end{equation}}

where $pixel_{mut}$ is the number of pixels modified on the image $I$ compared to $I_{\mathit{ori}}$, and $pixel_{total}$ denotes the total pixel count within the bounding boxes of objects in the original image $I_{\mathit{ori}}$ as identified by the MUT. Lowering $M_2$ reduces the number of perturbations in $I$  compared to $I_{\mathit{ori}}$. 


(3)~$M_3$ calculates the Euclidean distance between the original image $I_{\mathit{ori}}$ and the generated image $I$. This distance is normalized by the maximum Euclidean distance to a uniform image $I_{uni}$, either completely black or white, depending on which yields the larger $||\cdot||2$ with $I_{ori}$:

{\small\begin{equation}\label{objective_3_equation}
    M_3(I) = \frac{\| I_{ori} - I \|_2}{\| I_{uni} \|_2}
\end{equation}}
Lowering $M_3$ reduces the distance between $I$  and $I_{\mathit{ori}}$.

All the above three metrics are normalized so as not to overpower one another. The fitness function  is then defined as the weighted sum of the above three metrics: 
\begin{equation}\label{fitness_equation}
{ fitness (I) =w_1 \cdot M_1(I) + w_2 \cdot M_2 (I) + w_3 \cdot M_3 (I)}
\end{equation}

We initialize the weights based on these guidelines: we set $w_2$ and $w_3$ to high initial values to ensure minimal perturbations and to keep the generated images close to the original image. In contrast, we assign a low initial value to $w_1$ to gradually guide individuals toward attack solutions, i.e., images with undetectable objects, without overshadowing the other two metrics aimed at minimizing perturbations.

We use a plateau-based adaptation technique to dynamically adjust the weights (see lines~7-9 of Algorithm~\ref{alg:tmevo}). A plateau is identified when there is no improvement in the fitness score for $G_p$ consecutive generations, where $G_p$ is a parameter of our algorithm. The weights are assumed to be normalized, ranging between $0$ and $1$. Upon detecting a plateau, we adjust the weights: $w_1$ is increased by 5\%, while $w_2$ and $w_3$ are decreased by 5\%.  By incrementing $w_1$, we steer the fitness towards generating images that yield lower confidence scores by the MUT, i.e., are more likely to be an attack. Simultaneously, reducing $w_2$ and $w_3$ encourages the algorithm to accept slightly noisier attacks, which can enable the discovery of successful adversarial examples. Our plateau-based adaptation guides the attack generation to converge towards an equilibrium that is optimal for finding adversarial examples while keeping noise and degree of perturbation at a minimum.


\subsection{Mutation Noise Reduction}

TM-EVO uses the adaptive mutation operator from EvoAttack ~\cite{chan2022evoattack} as the genetic mutation operator (see Line 19 of Algorithm~\ref{alg:tmevo}). While this mutation operator  is optimized and adaptive, based on our experience, it may still introduce a sub-optimal amount of noise into the mutated images. We address this issue by applying an adaptive noise reduction operator  to the mutated images (see Line 20 of Algorithm~\ref{alg:tmevo}). This operator aims to reduce perturbations previously applied to a mutated image by a probability $\bar{\rho}$ which is a parameter of Algorithm~\ref{alg:tmevo}. The probability $\bar{\rho}$ is dynamically adjusted using our plateau-based adaptation.


Given an individual $I$, the noise reduction operator first selects, with the probability $\bar{\rho}$, a subset $b$ of the mutated pixels from $I$. The pixels in the selected subset $b$ in $I$ are, then,  replaced by their corresponding pixels from the original image. 
Let $\hat{I}$ represent the new image resulting from the noise removal process from $I$.  We check $\hat{I}$ post-noise reduction to ensure that $M_1(\hat{I})$, i.e., the metric assessing attacks, remains unchanged, while $M_2(\hat{I})$ and $M_3(\hat{I})$, i.e., the metrics assessing noise, are reduced (improved). If $\hat{I}$ does not meet these criteria, we omit it and keep using $I$. Otherwise, we replace $I$ with $\hat{I}$. The algorithm continuously adjusts $\bar{\rho}$ upon reaching a plateau, reducing it by $2$\%. 



\section{Empirical Evaluation}

In this section, we empirically assess TM-EVO through the following research question: 

\emph{RQ. How effective is TM-EVO in generating attack images with minimal noise compared to the original images? How efficient is TM-EVO in generating attacks?}

\textbf{Study subjects.} We use DETR~\cite{carion2020end} and Faster R-CNN~\cite{ren2015faster} as object-detection MUT.  
Faster R-CNN is based on a convolutional neural network (CNN) that has been applied to real-time applications like autonomous driving~\cite{ren2015faster}. DETR  uses a CNN backbone for feature extraction, and further, incorporates a transformer encoder-decoder architecture to detect objects~\cite{carion2020end}.

\textbf{Datasets.} We use the COCO~\cite{lin2015microsoft} and KITTI~\cite{geiger2013vision} datasets. The COCO dataset is widely used in computer vision, it contains diverse images annotated with object instances across multiple categories, e.g., animals and objects. The KITTI dataset includes images related to autonomous driving cars. The images in KITTI  are captured from sensors on cars, such as LiDAR and cameras, in urban environments. KITTI provides a real-world benchmark for testing the effectiveness of TM-EVO on object-detection models for autonomous driving.

\textbf{Parameters.} 
Table~\ref{tab:parameters} shows the parameters we used for TM-EVO experiments. Following earlier studies' parameter settings~\cite{chan2022evoattack} and our preliminary assessments, we set the population size $N$ to $32$, the maximum number of generations $G$ to $400$, the mutation rate $\rho$ to $0.024$, and the degree of perturbation $\delta$ to $0.4$. TM-EVO stops when either $400$ generations are reached or no objects with a confidence score above $0.9$ are detected. The initial weights are set based on the guidelines discussed in Section~\ref{subsec:fit}. We set the plateau-based adaptation parameter ($G_p$) to $10$ and the noise reduction rate ($\bar{\rho}$) to $0.3$. Based on our preliminary experiments, these settings for $G_p$ and $\bar{\rho}$ achieve a balance, producing minimal noise without significantly extending run time.




\begin{table}[t]
\begin{center}
\caption{Parameters of TM-EVO}
\label{tab:parameters}
\scalebox{0.7}{

\begin{tabular}{|l|l|l|l|}
\hline
Population size (N)     & $32$      & Maximum \# of generations ($G$)   & $400$   \\ 
\hline 
Mutation rate ($\rho$)  & $0.02$    & Degree of perturbation ($\delta$) & $0.4$   \\ 
\hline
Noise reduction rate ($\bar{\rho})$
                        & $0.3$      & \# of generation for adaptive plateau ($G_p$)                                                            & $10$  \\ 
\hline
Initial weights ($w_1$, $w_2$, $w_3$) 
                        &  $(0.1, 0.9,  0.9)$  
                                        &                                 &       \\ 
 \hline
\end{tabular}}
\end{center}
\vspace*{-.4cm}
\end{table}

\textbf{Experiment setup.} To address our research question, we compare TM-EVO with EvoAttack (hereafter referred to as EVO). We randomly select five images from the COCO dataset and five images from the KITTI dataset. TM-EVO and EVO are then applied to both DETR and Faster R-CNN using these images to generate adversarial attacks. Each run of each algorithm is repeated five times for both object-detection models to account for randomness. 

\textbf{Metrics.} The performance of TM-EVO and EVO is evaluated using three key metrics: L0 and L2 norms, as well as run time.  The L0 norm measures perturbation sparsity by counting the number of altered pixels, indicating the extent of visible changes in the generated attack images. The L2 norm calculates the Euclidean distance between the original and the generated attack images, reflecting the subtlety of modifications. Lower values in L0 and L2 metrics indicate less noticeable alterations in the generated attack images. Run time is the time taken by each algorithm to generate an adversarial attack. 

\begin{figure}
    \includegraphics[width=122pt, height=120pt]{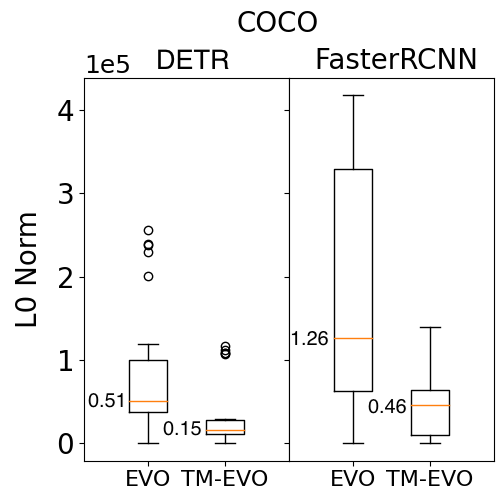}
    \includegraphics[width=122pt, height=120pt]{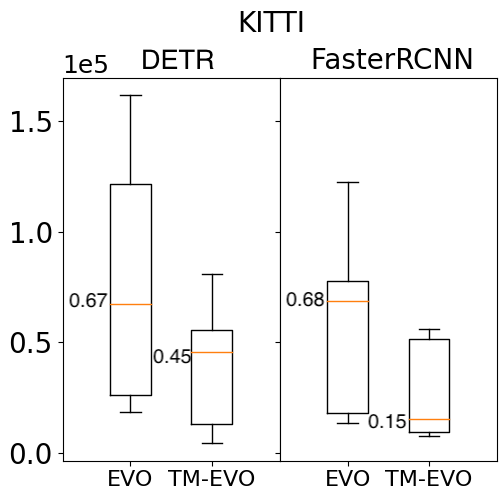}\\
    \includegraphics[width=122pt, height=120pt]{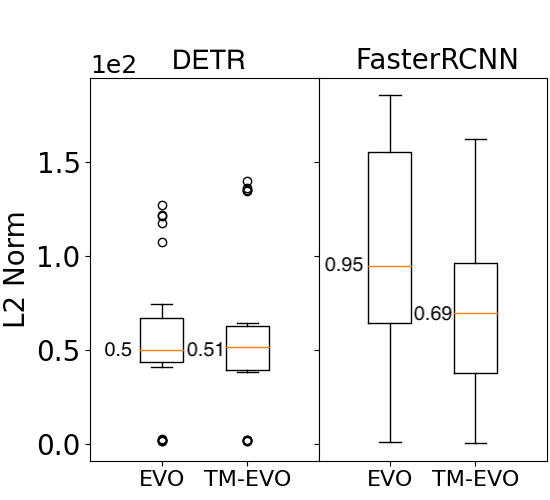}
    \includegraphics[width=122pt, height=120pt]{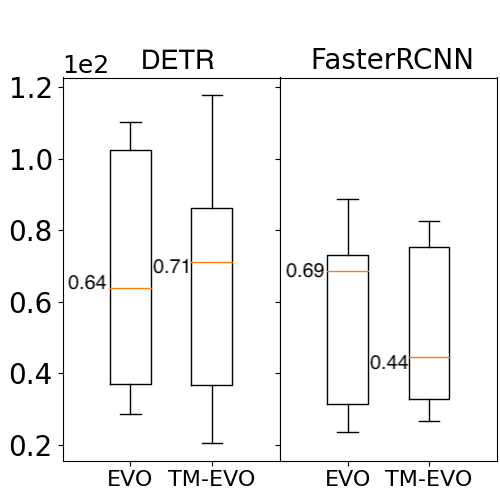}\\
    \includegraphics[width=122pt, height=120pt]{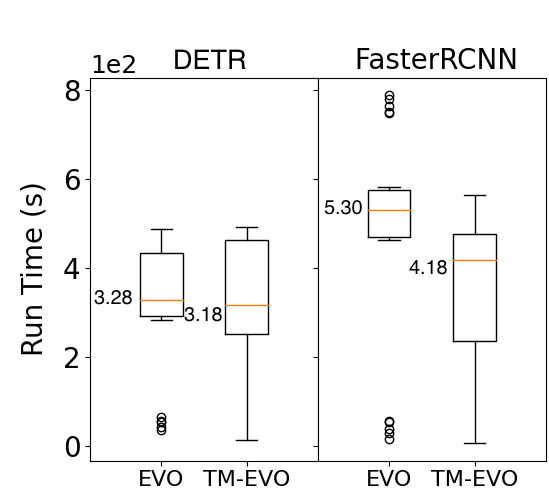}
    \includegraphics[width=122pt, height=120pt]{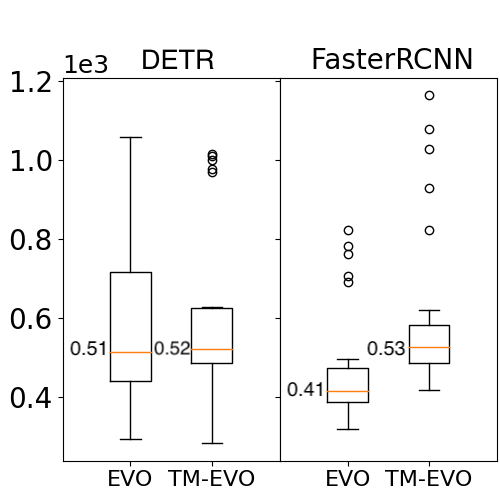}
    \vspace*{-.2cm}
  \caption{Comparing TM-EVO and EVO algorithms: (a) L0 norm results, (b) L2 norm results, and (c) run time results.}
  \label{fig:plots}
   \vspace*{-.3cm}
\end{figure}

\textbf{Results.} Figure~\ref{fig:plots} presents the results comparing the performance of TM-EVO and EVO. Note that both TM-EVO and EVO successfully found an attack for all the input images in all runs.
Hence, in Figure~\ref{fig:plots}, we focus on L0 and L2 norm and run time results to compare TM-EVO and EVO in efficiently generating stealthy adversarial attacks. 

As the figure shows, on average, TM-EVO reduces noise in the generated attacks by 60.8\% compared to EVO, as measured by the L0 norm metric. We also compared the L0 norm results using the non-parametric pairwise Wilcoxon rank sum test. The results indicate that TM-EVO yields statistically better L0 norm values compared to EVO, with a confidence level of 95\%.  The L2 norm results from TM-EVO and EVO are almost identical for DETR on the COCO dataset. However, TM-EVO is, on average, $10.2$\% worse than EVO for DETR on the KITTI dataset. In contrast, for Faster R-CNN, TM-EVO improves L2 norm results by an average of $30.9$\% over EVO for both datasets. Considering all datasets and models, TM-EVO yields slightly better L2 norm results than EVO. Regarding run time, neither approach consistently outperforms the other. TM-EVO and EVO exhibit similar run times for DETR on KITTI and COCO datasets. For Faster R-CNN, TM-EVO outperforms EVO on the COCO dataset by an average of $111.9$s, while it lags on the KITTI dataset by an average of $112.4$s.

\emph{In summary}, TM-EVO outperforms EVO in reducing noise in the generated attacks as measured by L0 and L2 norm metrics across two object-detection models. This improvement is achieved without a notable rise in average run time. 

Finally, comparing DETR and Faster R-CNN yields interesting contrasts: In the case of DETR, the L0 and L2 norm results from EVO and TM-EVO are more similar compared to those for Faster R-CNN, where TM-EVO significantly outperforms EVO. In particular, with DETR, TM-EVO outperforms EVO in terms of L0 only. However, for Faster R-CNN, TM-EVO outperforms EVO in both noise-measurement metrics.  This implies a limited potential for minimizing the degree of image perturbation for attacks generated for DETR, which uses a Transformer architecture, 
compared to those generated for Faster R-CNN's CNN-based framework.


\section{Conclusion}

We show the potential of multi-metric evolutionary search in achieving adversarial attacks with minimal noise, and leverage its adaptability to finely tune the generation of adversarial attacks and the required noise levels. 
In future work, we aim to explore the validity of the generated attacks to see if they are detectable by humans or anomaly detectors~\cite{RiccioT23}. We also plan to compare  TM-EVO with multi-objective search and the performance of the attacks generated by TM-EVO in online testing~\cite{HaqSNB20}. Our replication package is available online~\cite{github}. 


\section{References}
\printbibliography[heading=none]

\end{document}